%% file: 0PaperID100.tex
\documentclass[sigconf,pbalance]{acmart}

\AtBeginDocument{%
  \providecommand\BibTeX{{%
    \normalfont B\kern-0.5em{\scshape i\kern-0.25em b}\kern-0.8em\TeX}}}

\copyrightyear{2026}
\acmYear{2026}
\setcopyright{cc}
\setcctype{by-nc-nd}
\acmConference[SIGIR '26]{Proceedings of the 49th International ACM SIGIR Conference on Research and Development in Information Retrieval}{July 20--24, 2026}{Melbourne, VIC, Australia}
\acmBooktitle{Proceedings of the 49th International ACM SIGIR Conference on Research and Development in Information Retrieval (SIGIR '26), July 20--24, 2026, Melbourne, VIC, Australia}
\acmDOI{10.1145/3805712.3808609}
\acmISBN{979-8-4007-2599-9/2026/07}

\usepackage{enumitem}
\usepackage{amsmath}
\usepackage{multirow}
\usepackage{booktabs}

\usepackage{amsfonts}
\usepackage{makecell}
\usepackage{graphicx}
\usepackage{subcaption}
\usepackage{cleveref}
\usepackage{mdframed}
\usepackage{fvextra}   %
\usepackage{pifont} %
\usepackage{dirtree}
\usepackage{tcolorbox}  %
\usepackage{tikz}
\tcbuselibrary{breakable}   %
\usepackage{xcolor}
\usepackage[table]{xcolor}
\usepackage{marvosym}

\begin{document}

\title{Personalized Deep Research: A User-Centric Framework, Dataset, and Hybrid Evaluation for  Knowledge Discovery}

\author{Xiaopeng Li}
\affiliation{%
  \institution{City University of Hong Kong}
  \city{Hong Kong}
  \country{China}
}
\email{xiaopli2-c@my.cityu.edu.hk}

\author{Wenlin Zhang}
\affiliation{%
  \institution{City University of Hong Kong}
  \city{Hong Kong}
  \country{China}
}
\email{wl.z@my.cityu.edu.hk}

\author{Yingyi Zhang}
\affiliation{%
  \institution{City University of Hong Kong}
  \city{Hong Kong}
  \country{China}
}
\email{yingyizhang@mail.dlut.edu.cn}

\author{Pengyue Jia}
\affiliation{%
  \institution{City University of Hong Kong}
  \city{Hong Kong}
  \country{China}
}
\email{jia.pengyue@my.cityu.edu.hk}

\author{Yejing Wang}
\affiliation{%
  \institution{City University of Hong Kong}
  \city{Hong Kong}
  \country{China}
}
\email{yejing.wang@my.cityu.edu.hk}

\author{Yichao Wang\footnotemark[1]}
\affiliation{%
  \institution{Huawei Technologies Ltd.}
  \city{Shenzhen}
  \country{China}
}
\email{wangyichao5@huawei.com}

\author{Yong Liu}
\affiliation{%
  \institution{Huawei Technologies Ltd.}
  \city{Shenzhen}
  \country{China}
}
\email{liu.yong6@huawei.com}

\author{Huifeng Guo}
\affiliation{%
  \institution{Huawei Technologies Ltd.}
  \city{Singapore}
  \country{Singapore}
}
\email{huifeng.guo@huawei.com}

\author{Xiangyu Zhao}
\authornote{Corresponding authors.}
\affiliation{%
  \institution{City University of Hong Kong}
  \city{Hong Kong}
  \country{China}
}
\email{xianzhao@cityu.edu.hk}

\renewcommand{\shortauthors}{Xiaopeng Li et al.}

\begin{abstract}
Deep Research agents driven by LLMs have automated the scholarly discovery pipeline, from planning and query formulation to iterative web exploration. Yet they remain constrained by a static, ``one-size-fits-all'' retrieval paradigm. Current systems fail to adaptively adjust the depth and breadth of exploration based on the user's existing expertise or latent interests, frequently resulting in reports that are either redundant for experts or overly dense for novices. To address this, we introduce Personalized Deep Research (PDR), a framework that integrates dynamic user context into the core retrieval-reasoning loop. Rather than treating personalization as a post-hoc formatting step, PDR unifies user profile modeling with iterative query development, dual-stage (private/public) retrieval, and context-aware synthesis. This allows the system to autonomously align research sub-goals with user intent and optimize the stopping criteria for evidence collection. To facilitate benchmarking, we release the PDR Dataset, covering four realistic user tasks, and propose a hybrid evaluation framework combining lexical metrics with LLM-based judgments to assess factual accuracy and personalization alignment. Experimental results against commercial baselines demonstrate that PDR significantly improves retrieval utility and report relevance, effectively bridging the gap between generic information retrieval and personalized knowledge acquisition. The resource is available to the public at~\url{https://github.com/Applied-Machine-Learning-Lab/SIGIR2026_PDR}.
\end{abstract}



\maketitle
\input{1Introduction}
\input{3Methodology}
\input{4Dataset}
\input{5Evaluation}
\input{6Results}

\input{2Relatedwork}
\input{7Conclusion}

\bibliographystyle{ACM-Reference-Format}
\bibliography{bibfile}

\end{document}

%% file: 1Introduction.tex
\section{Introduction}
The rapid advancement of artificial intelligence has revolutionized the pipeline of knowledge discovery in both academic and industrial settings. Conventional knowledge-intensive research tasks require experts to formulate research questions, conduct extensive literature reviews, analyze findings, and synthesize comprehensive research reports. However, recent developments in \textbf{Deep Research} frameworks~\cite{li2025webthinker, zheng2025deepresearcher, schmidgall2025agent, tang2025ai, zhang2025survey_dr}, such as \textit{OpenAI Deep Research}~\cite{openai_deep_research_2025} and \textit{Perplexity Deep Research}~\cite{perplexity_deep_research_2025}, have fundamentally transformed this workflow. By integrating large language models with advanced reasoning capabilities and adaptive retrieval systems, these frameworks autonomously orchestrate iterative retrieval-reasoning cycles. This process significantly reduces research duration while delivering high-quality, evidence-based outputs. These efforts align with the broader paradigm shift towards AI-driven search and information seeking~\cite{li2025aisearch, zhao2018drl}.

Existing Deep Research frameworks typically operate through three tightly coupled stages: \textbf{(1) Planning}, in which the agent decomposes the overarching research question into an ordered sequence of sub-goals to construct a task-aware roadmap prior to execution; \textbf{(2) Searching}, where the agent dynamically interacts with external environments by formulating context-sensitive queries and performing iterative retrieval. This stage addresses evolving information needs through rigorous noise filtering to achieve coverage deeper than that of standard retrieval-augmented generation pipelines; and \textbf{(3) Report Generation}, where the agent synthesizes curated evidence into a structured document. This process involves selecting salient passages and organizing discourse to produce output that approximates human-authored research reports. Empirically, these systems have demonstrated proficiency in two primary scenarios: (1) addressing complex benchmarks such as HLE~\cite{phan2025humanity}, GAIA~\cite{mialon2024gaia}, and SimpleQA~\cite{wei2024measuring}; and (2) generating comprehensive reports in minutes that rival the quality of expert analysis~\cite{gemini_deep_research_2024}.

\begin{figure}
    \centering
    \includegraphics[width=0.95\linewidth]{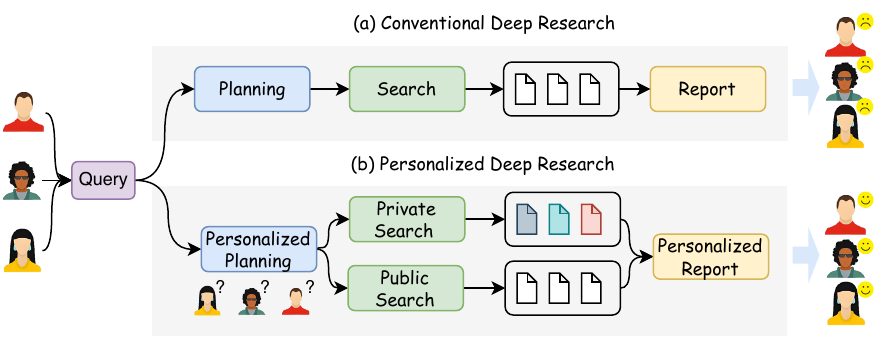}
    \vspace{-2mm}
    \caption{Comparison between Conventional and Personalized Deep Research pipelines.
(a) The conventional approach lacks user context and leads to generic outputs.
(b) Personalized Deep Research leverages user-specific knowledge for tailored and satisfactory results. }
\vspace{-2mm}
    \label{fig:motivation}
\end{figure}

Despite these achievements, current Deep Research solutions remain fundamentally {\textbf{``one-size-fits-all''}} (see Figure~\ref{fig:motivation}). These pipelines are primarily optimized on broad, domain-agnostic corpora and often fail to account for the specific preferences and contextual requirements of users. Consequently, the generated reports frequently do not meet distinct needs, such as the stylistic conventions of an academic writer, the formatting guidelines of a consultant, or the granularity required by a policy analyst. This lack of personalization presents three critical challenges: (1) the effective integration and exploitation of personal information within the Deep Research workflow; (2) the absence of publicly available datasets that support personalized research behaviors; and (3) the limitation of existing evaluation metrics in comprehensively assessing alignment with user preferences.

To address these limitations, we present a streamlined Personalized Deep Research (PDR) pipeline in this paper: \textbf{(1) Personalized Deep Research (PDR) Framework.} We extend the canonical three-stage workflow by integrating user-specific information throughout the process via four dedicated personalization modules: \ding{182} \emph{Profile Extraction}, which serves as the foundation for personalization by structuring historical documents, interactions, and metadata into a comprehensive user profile containing demographics, traits, and preferences; \ding{183} \emph{Personalized Question Development}, which facilitates a fine-grained understanding of user intent by tailoring research sub-goals to individual profiles; \ding{184} \emph{Dynamic Dual-Stage Retrieval}, which accesses both external knowledge bases and private user repositories through an iterative loop to ensure the retrieval of both precise factual knowledge and relevant personalized context; and \ding{185} \emph{Personalized Report Generation}, which synthesizes user profiles, original queries, and retrieved evidence to produce reports aligned with user needs.
\textbf{(2) PDR Dataset.} We release the first dataset dedicated to personalized Deep Research, encompassing four realistic scenarios: personalized abstract generation, personalized topic writing, personalized report generation, and personalized speech script generation. Unlike synthetic alternatives, our data is derived from authentic real-world scenarios and rigorously anonymized to ensure alignment with practical applications while maintaining data security.
\textbf{(3) Comprehensive Evaluation Protocol PDR-Eval.} We propose a hybrid evaluation framework that combines lexical metrics with an LLM-as-Judge approach. This protocol comprehensively assesses system performance across multiple dimensions, focusing on both the factual quality of the content and the degree of personalization.

In summary, the contributions of this work are as follows:

\begin{itemize}[leftmargin=*]
\item  We propose a pioneering Personalized Deep Research framework that seamlessly integrates user-specific preferences into the deep research workflow, thereby significantly enhancing both user experience and output relevance.
\item To address the lack of dedicated datasets, we construct a comprehensive benchmark based on real-world scenarios. This benchmark encompasses four representative task categories to facilitate future research. Furthermore, we design a novel evaluation protocol tailored to assess both factual accuracy and the quality of personalization. 
\item Extensive experiments demonstrate that our framework significantly outperforms iterative RAG and existing industrial systems, delivering superior and user-aligned research outputs.

\end{itemize}

%% file: 3Methodology.tex
\begin{figure*}
    \centering
    \includegraphics[width=\linewidth]{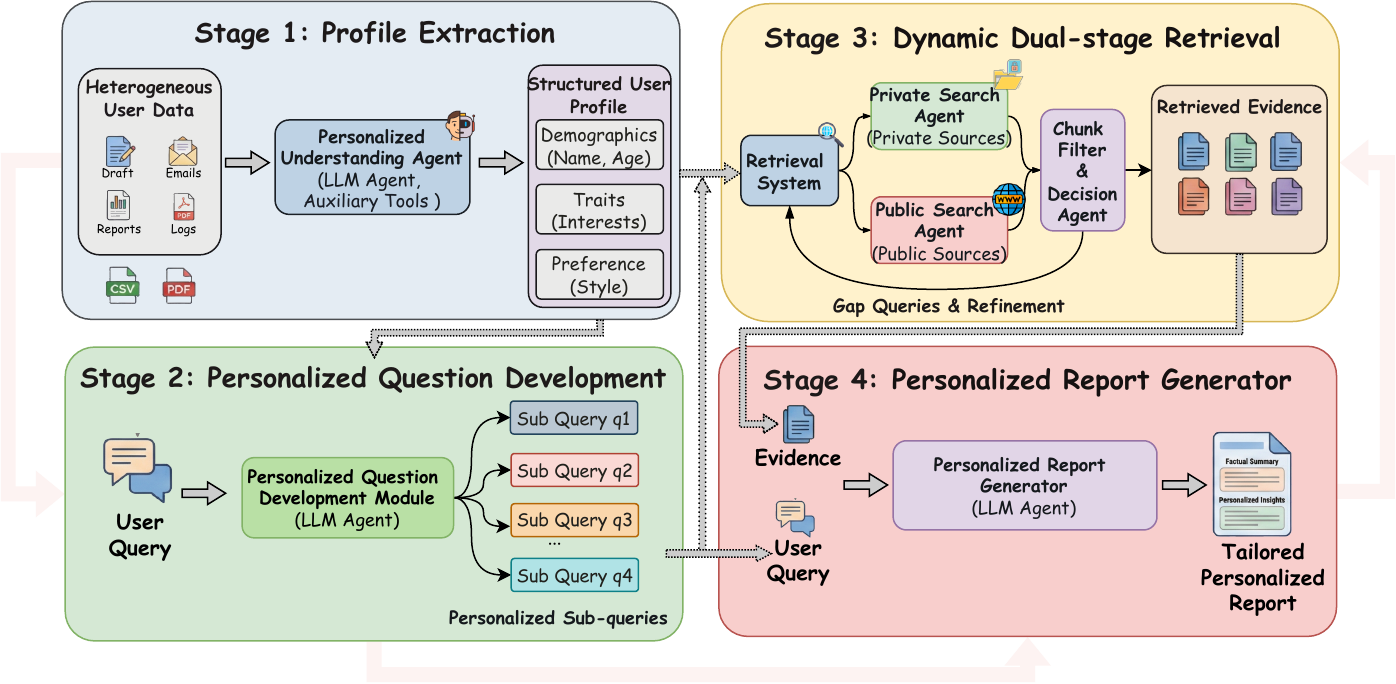}
    \caption{ Overview of the Personalized Deep Research (PDR) framework. It consists of four core stages: (i) profile extraction from user data, (ii) personalized question
    development, (iii) dynamic dual-stage retrieval integrating private and external sources, and (iv) personalized report generation.}
    \label{fig:framework}
\end{figure*}

\section{Framework of Personalized Deep Research}
\label{sec:Framework}
\subsection{Overview}
As illustrated in Figure \ref{fig:framework}, our Personalized Deep Research (PDR) framework operates through a four-stage pipeline. First, the Profile Extraction module transforms heterogeneous user data into structured profiles, capturing demographics, behavioral traits, and preferences. Second, Personalized Question Development contextualizes user input with these profiles, decomposing queries to resolve latent user intent. Third, Dynamic Two-stage Retrieval accesses both private and public corpora; a decision agent iteratively assesses findings and issues targeted 'gap queries' to address knowledge deficiencies. Finally, the Personalized Report Generator synthesizes retrieved content into reports that align with the user's profile while ensuring factual accuracy.

\subsection{Profile Extraction}
Traditional deep-research pipelines typically operate under a ``one-size-fits-all'' paradigm, failing to adapt to individual user preferences due to insufficient personalization mechanisms. A primary challenge in addressing this is data heterogeneity: user information is fragmented across diverse sources (e.g., drafts, emails, browsing logs) and formats (CSV, PDF, Markdown). We address this by implementing a Personalized Understanding Agent powered by a reasoning-oriented LLM. This agent dynamically orchestrates specialized auxiliary tools to parse, interpret, and synthesize content from these varied sources. Formally, we model the profile generation process as:
\begin{equation}
P(u) = f_{\text{LLM}}(\cup_{i=1}^{n} D_i; \mathcal{T})
\end{equation}
where $P(u)$ represents the personalized profile for user $u$, $D_i$ denotes raw data from source $i$, and $\mathcal{T}$ represents the set of auxiliary tools used for processing diverse formats. Additionally, we establish a standardized schema to capture essential attributes, including demographics, learning interests, response preferences, and interaction tendencies. This structured profile provides a stable foundation for personalization, ensuring consistent utility across all subsequent stages of the deep-research pipeline.

\subsection{Personalized Question Development}
Effective deep research relies on robust question development, necessitating the iterative generation of targeted, context-aware sub-queries rather than static keyword matching. However, existing pipelines fail to integrate user personalization during this critical phase. Even state-of-the-art systems, such as OpenAI Deep Research~\cite{openai_deep_research_2025} and Gemini Deep Research~\cite{gemini_deep_research_2024}, rely on reactive, manual clarification loops rather than automated intent modeling, often decoupling retrieval from specific user needs. To address this, we introduce a Personalized Question Development module that synthesizes user profiles with input queries to automate intent-aware decomposition. We formalize this process as:
\begin{equation}
Q_{sub} = f_{\text{LLM}}(Q_{\text{original}}, \mathcal{P}(u)) = \{q_1, q_2, ..., q_k\}
\end{equation}
where $Q_{\text{original}}$ is the initial query, $\mathcal{P}(u)$ denotes the personalization profile for user $u$, and $Q_{sub}$ is the resulting set of $k$ optimized sub-queries. These sub-queries are dispatched in parallel, ensuring that subsequent retrieval is strictly aligned with user-specific constraints and preferences without requiring manual intervention.

\subsection{Dynamic Dual-stage Retrieval}

Deep research workflows necessitate the extraction of precise evidence from heterogeneous corpora. This requirement is particularly critical in personalized generation tasks involving private documents, where relevant information is often sparsely distributed. To address this challenge, our framework integrates user-specific context with general domain knowledge; for instance, the system aligns a work plan with historical reports of the user while simultaneously sourcing external data. We implement this approach via a dual-agent retrieval paradigm that unifies the mining of private documents with the search for public knowledge:
\begin{equation}
\mathcal{R}(q, \mathcal{P}(u)) = \mathcal{R}_{\text{internal}}(q, \mathcal{D}_{\text{private}}) \cup \mathcal{R}_{\text{external}}(q, \mathcal{D}_{\text{public}})
\end{equation}
where $\mathcal{R}_{\text{internal}}$ and $\mathcal{R}_{\text{external}}$ denote retrieval functions over private ($\mathcal{D}_{\text{private}}$) and public ($\mathcal{D}_{\text{public}}$) repositories, respectively. To ensure adaptability, the framework incorporates three distinct mechanisms: (1) a chunk-filtering agent that eliminates irrelevant content to enhance precision; (2) a decision agent that dynamically determines the necessity of external retrieval or additional search iterations; and (3) a query-evolution mechanism that iteratively refines queries based on intermediate results. This architecture efficiently balances personalized context with comprehensive external coverage.

\subsection{Personalized Report Generation}
The Personalized Report Generation module serves as the final stage of our PDR pipeline, synthesizing fragmented findings from parallel sub-queries into coherent, evidence-rich documents. While existing systems struggle to balance factual integrity with user-specific communication styles, our approach resolves this tension through a dynamic structure-control mechanism. This mechanism adapts section ordering, content depth, and tone based on the user's profile, ensuring outputs align with both verified evidence and individual preferences. The generation process follows a systematic three-step workflow: (1) aggregation of sub-query results and retrieved segments; (2) integration of these materials with the personalization profile to establish a comprehensive context; and (3) final report synthesis via a Large Language Model (LLM) employing chain-of-thought reasoning for style adaptation:
\begin{equation}
\mathcal{T}_{\text{final}} =
f_{\text{gen}}\left(
\bigcup_{q \in \mathcal{Q}} \mathcal{R}(q, \mathcal{P}(u)), \mathcal{P}(u), \mathcal{Q}
\right)
\end{equation}
where $\mathcal{T}_{\text{final}}$ represents the personalized report, $\bigcup_{q \in Q} \mathcal{R}$ denotes the union of retrieved results across all sub-queries $Q$, and $\mathcal{P}(u)$ is the user personalization profile. This formulation ensures that the final output maximizes analytical utility while strictly adhering to the user's established communication patterns.

%% file: 4Dataset.tex
\begin{figure}
    \centering
    \includegraphics[width=0.95\linewidth]{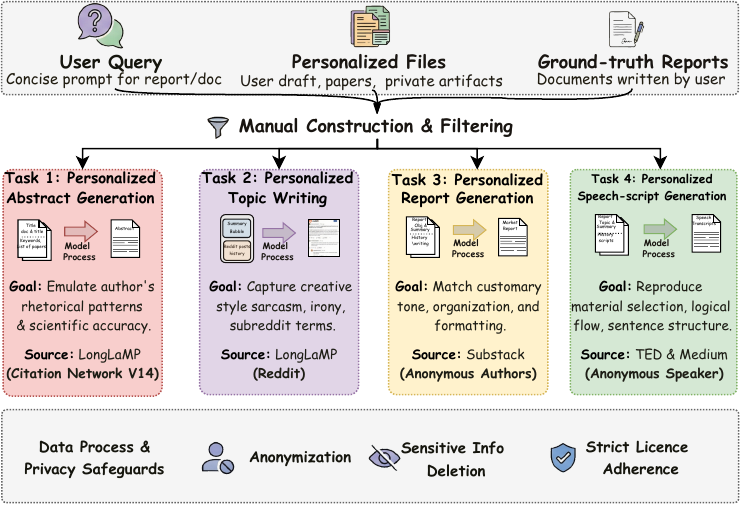}
    \caption{Dataset Construction Pipeline for PDR.}
    \label{fig:dataset}
\end{figure}

\section{Dataset}

Standard Deep Research evaluations typically rely on challenging Question Answering (QA) tasks, such as Simple QA~\cite{wei2024measuring} and HLE~\cite{phan2025humanity}, alongside tool-use benchmarks like GAIA~\cite{mialon2024gaia}. Recently, dedicated Deep Research benchmarks have been developed to build datasets of PhD-level research tasks~\cite{xu2025researcherbench} and real-world scientific-scenario question sets~\cite{du2025deepresearch}. These benchmarks aim to assess agents' ability to compose extended research reports. Our work diverges by focusing on personalized scenarios. In personalized contexts, the primary concern is whether Deep Research systems can generate reports aligning closely with individual preferences and established user writing patterns. Despite this practical demand, no public dataset currently supports personalized Deep Research evaluation. To operationalize the pipeline proposed in Section~\ref{sec:Framework}, a suitable dataset must contain three essential elements:

\begin{itemize}[leftmargin=*]
    \item \textbf{User queries:} Typically a concise prompt summarizing the desired report or document.
    \item \textbf{Personalized files:} User drafts, notes, papers, and other private artifacts.
    \item \textbf{Ground-truth reports:} Documents actually written by the user.
\end{itemize}

Since no existing corpus satisfies all three requirements, we manually constructed a comprehensive dataset by filtering public resources and collecting additional documents. The resulting benchmark comprises four distinct task types designed to evaluate different aspects of personalized content generation.

\subsection{Task 1: Personalized Abstract Generation}

This task requires producing detailed abstracts that faithfully represent a paper's contribution while emulating the author's rhetorical patterns. The input consists of the paper title, selected keywords from the original abstract, and the author's prior publications. The expected output is a rewritten abstract maintaining scientific accuracy while reflecting the author's personal writing style. We curated this subset by filtering the LongLaMP~\cite{kumar2024longlamp} corpus, derived from the Citation Network Dataset (V14)~\cite{tang2008arnetminer}. We retained twenty users whose abstracts exceed 2,000 characters to ensure the task remains knowledge-intensive and provides sufficient content for meaningful personalization assessment.

\subsection{Task 2: Personalized Topic Writing}

This task focuses on generating complete Reddit posts reflecting the author's creative style, including sarcasm, irony, and subreddit-specific terminology. The model receives a concise summary of the prospective post along with the user's previous Reddit submissions and must produce a fully developed post capturing the author's distinctive voice and communication patterns. Data originate from LongLaMP~\cite{kumar2024longlamp}, derived from the Reddit TL;DR corpus~\cite{volske2017tl}. We filtered twenty users with target texts longer than 5,000 characters to preserve task difficulty and ensure adequate complexity for evaluating personalization capabilities.

\subsection{Task 3: Personalized Report Generation}

This task requires drafting comprehensive reports, such as market analyses, investigative studies, or annual reviews, matching the writer's customary tone, content organization, and formatting conventions. Inputs include a report objective summary plus the author's historical writings. The desired output is a complete report demonstrating both factual accuracy and stylistic consistency with the author's established patterns. We assembled this dataset by collecting Substack~\cite{substack2025} content from five prolific authors. We strictly adhered to data usage licenses and implemented comprehensive privacy protection measures. All personal identifiers were anonymized, sensitive information removed, and rigorous privacy safeguards established before including the material in our benchmark.

\subsection{Task 4: Personalized Speech-Script Generation}

This task targets producing complete speeches that reproduce the speaker's characteristic material selection, logical flow, and sentence structure. The system receives a topic summary together with the speaker's prior scripts and must output a complete transcript maintaining both topical relevance and authentic stylistic representation. We sourced publicly available talks from TED~\cite{TEDWebsite} and essays from Medium~\cite{MediumWebsite}, selecting five speakers for inclusion. We applied identical privacy safeguards and strictly followed data usage licenses, implementing comprehensive privacy protection measures, including name anonymization and sensitive data deletion, before incorporating the content into our dataset.

%% file: 5Evaluation.tex
\begin{figure}
    \centering
    \includegraphics[width=0.95\linewidth]{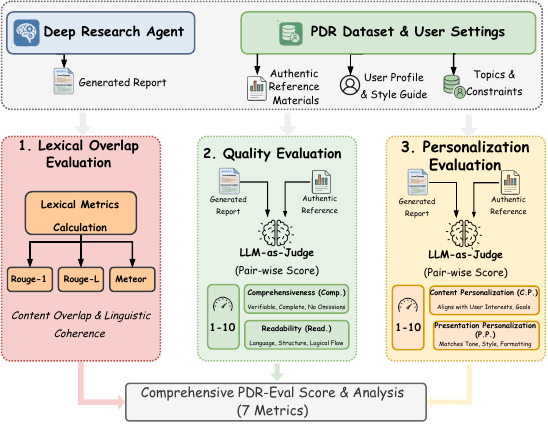}
    \caption{Overview of the PDR-Eval Framework for Deep Research. Incorporating Lexical Overlap, Quality Evaluation, and Personalization Evaluation.}
    \label{fig:evaluation}
\end{figure}

\section{Evaluation}

Evaluating deep research agents presents a significant challenge due to their complex internal architectures, which complicate process-based assessment. Consequently, evaluation methodologies must prioritize the quality of the final generated reports. While deep research evaluation is nascent, early frameworks like DeepResearch Bench~\cite{du2025deepresearch} and ResearcherBench~\cite{xu2025researcherbench} have introduced PhD-level tasks and rubric-based assessments for report quality and retrieval accuracy. However, in our Personalized Deep Research (PDR) setting, evaluation must extend beyond factual correctness to incorporate personalization metrics. Leveraging our dataset's authentic abstracts, reports, and speech scripts, we propose \textbf{PDR-Eval}, a framework assessing performance across three dimensions: Lexical Overlap, Quality, and Personalization.

\noindent \paragraph{\textbf{Lexical Overlap Evaluation}}
We employ ROUGE-1, ROUGE-L~\cite{lin-2004-rouge}, and METEOR~\cite{banerjee2005meteor} to quantify lexical similarity between generated and reference documents. These metrics provide an objective comparison of content overlap and linguistic coherence.

\paragraph{\textbf{Quality Evaluation}}
We utilize an LLM-as-Judge approach to assess \textit{Comprehensiveness} and \textit{Readability}. Adopting a pair-wise scoring strategy inspired by DeepResearch Bench~\cite{du2025deepresearch}, we simultaneously submit the generated report and the authentic reference document (serving as the gold standard) to the evaluator. Performance is rated on a 10-point scale:
\begin{itemize}[leftmargin=*]
    \item \textbf{Comprehensiveness (Comp.):} Measures the extent to which factual statements are verifiable, correct, and complete. Full scores require the inclusion of all essential sub-topics, data points, and contextual elements without material omissions.
    \item \textbf{Readability (Read.):} Assesses how easily the target audience can comprehend the report based on language, sequencing, and structure. Optimal readability requires syntax and vocabulary matched to audience proficiency, clear logical transitions, and a hierarchical structure facilitating rapid information retrieval.
\end{itemize}

\noindent\paragraph{\textbf{Personalization Evaluation}}
Similarly, we employ an LLM-as-Judge approach to evaluate \textit{Content} and \textit{Presentation Personalization}. Using the same pair-wise scoring method against authentic references, we define the metrics as follows:
\begin{itemize}[leftmargin=*]
    \item \textbf{Contextual Personalization (C. P.):} Measures the alignment of selected information (topics, examples, ordering) with explicit or inferred user interests and goals. Perfect personalization implies that every element maps directly to a user need, strictly excluding irrelevant content.
    \item \textbf{Presentation Personalization (P. P.):} Evaluates the conformity of tone, style, formatting, and media choices to user preferences or brand requirements. Full scores are awarded when the output matches specified templates and requires no post-production editing.
\end{itemize}

%% file: 6Results.tex
\section{Experiment}
In this section, we present detailed experimental validation results for our proposed framework. We compare our approach to various baselines and provide a comprehensive analysis.

\begin{table*}[t]
  \centering
  \small
  \renewcommand{\arraystretch}{0.95} %
  \setlength{\tabcolsep}{7.2pt} %
  \caption{Performance comparison on Task 1 \& 2 (Top) and Task 3 \& 4 (Bottom). Deep Research Agents show competitive performance in specific metrics, while \textbf{PDR (Ours)} maintains dominance in Personalization. The best-performing value is in \textbf{Bold}, and the second-best is marked with \underline{underline}}
  \label{tab:R1_performance_compact}
  
  \resizebox{\textwidth}{!}{
  \begin{tabular}{l|ccccccc|ccccccc}
    \toprule
    \multicolumn{1}{c|}{\multirow{2}{*}{\textbf{Methods}}} & \multicolumn{7}{c|}{\textbf{Task 1}} & \multicolumn{7}{c}{\textbf{Task 2}} \\
    \cmidrule(lr){2-8} \cmidrule(lr){9-15}
    & R-1 & R-L & Met. & Comp. & Read. & C.P. & P.P. & R-1 & R-L & Met. & Comp. & Read. & C.P. & P.P. \\
    \midrule
    \multicolumn{15}{l}{\textit{Non-personalized}} \\
    \quad Zero-shot         & 0.2996 & 0.1316 & 0.1202 & 4.55 & 6.93 & 4.81 & 7.78 & 0.1986 & 0.0860 & 0.0707 & 3.88 & 7.25 & 4.37 & 6.89 \\
    \quad +Search           & 0.2903 & 0.1420 & 0.1204 & 4.60 & 6.92 & 5.25 & 7.93 & 0.2176 & 0.0947 & 0.0974 & 3.89 & 7.30 & 4.86 & 7.14 \\
    \midrule
    \multicolumn{15}{l}{\textit{Deep Research Agents}} \\
    \quad Grok-DR           & \underline{0.3110} & 0.1410 & \underline{0.1240} & 5.40 & 7.50 & 5.00 & 7.10 & \textbf{0.3458} & \textbf{0.1233} & \textbf{0.2024} & \textbf{6.26} & \textbf{8.57} & \underline{6.18} & 7.65 \\
    \quad Perplexity-DR     & 0.3050 & 0.1380 & 0.1220 & 5.10 & 7.20 & 5.15 & 7.30 & 0.2343 & 0.0948 & 0.1155 & 4.87 & 7.65 & 6.04 & 7.20 \\
    \quad Gemini-DR         & \textbf{0.3485} & \textbf{0.1540} & \textbf{0.1310} & \textbf{6.10} & \textbf{7.95} & 5.20 & 7.05 & \underline{0.2788} & \underline{0.1024} & \underline{0.1234} & \underline{5.93} & \underline{8.25} & 5.35 & 6.76 \\
    \quad OpenAI-DR         & 0.3020 & 0.1395 & 0.1210 & 5.30 & 7.40 & 5.40 & 7.55 & 0.2286 & 0.0843 & 0.0844 & 5.10 & 7.84 & 5.58 & 7.82 \\
    \midrule
    \multicolumn{15}{l}{\textit{Personalized}} \\
    \quad Profile Prompting & 0.2962 & 0.1480 & 0.1209 & 5.28 & 7.43 & 5.38 & 8.42 & 0.2157 & 0.0905 & 0.0748 & 4.43 & 7.44 & 5.32 & 6.87 \\
    \quad Iterative RAG     & 0.3073 & 0.1522 & 0.1182 & 4.78 & 6.85 & \underline{6.04} & \underline{8.79} & 0.1619 & 0.0761 & 0.0579 & 4.48 & 7.45 & 5.78 & \underline{7.95} \\
    \quad \textbf{PDR (Ours)} & 0.3099 & \underline{0.1532} & 0.1211 & \underline{5.82} & \underline{7.71} & \textbf{7.82} & \textbf{9.51} & 0.2455 & 0.0971 & 0.0964 & 5.24 & 7.90 & \textbf{6.29} & \textbf{8.51} \\
    
    \bottomrule
    \toprule
    \multicolumn{1}{c|}{\multirow{2}{*}{\textbf{Methods}}} & \multicolumn{7}{c|}{\textbf{Task 3}} & \multicolumn{7}{c}{\textbf{Task 4}} \\
    \cmidrule(lr){2-8} \cmidrule(lr){9-15}
    & R-1 & R-L & Met. & Comp. & Read. & C.P. & P.P. & R-1 & R-L & Met. & Comp. & Read. & C.P. & P.P. \\
    \midrule
    \multicolumn{15}{l}{\textit{Non-personalized}} \\
    \quad Zero-shot         & 0.3378 & 0.1104 & 0.2768 & 8.30 & 8.08 & 3.78 & 6.09 & 0.3799 & 0.1190 & 0.2583 & 7.49 & 8.43 & 7.85 & 5.90 \\
    \quad +Search           & 0.3503 & 0.1230 & 0.2965 & 8.49 & 8.12 & 3.64 & 5.77 & 0.3812 & 0.1342 & 0.2746 & 7.73 & 8.58 & 7.98 & 6.02 \\
    \midrule
    \multicolumn{15}{l}{\textit{Deep Research Agents}} \\
    \quad Grok-DR           & 0.3650 & 0.1290 & 0.2950 & 8.80 & 8.35 & 4.10 & 6.40 & 0.3950 & 0.1380 & 0.2850 & 8.90 & 8.95 & 8.00 & 7.40 \\
    \quad Perplexity-DR     & \textbf{0.3920} & \textbf{0.1410} & \textbf{0.3210} & \textbf{9.20} & \textbf{8.65} & 4.25 & 6.55 & \textbf{0.4120} & \textbf{0.1450} & \textbf{0.2980} & 8.85 & 8.80 & 7.80 & 7.10 \\
    \quad Gemini-DR         & \underline{0.3710} & \underline{0.1320} & \underline{0.3050} & 8.95 & \underline{8.45} & 4.00 & 6.20 & \underline{0.4010} & \underline{0.1410} & \underline{0.2920} & \textbf{9.30} & \textbf{9.20} & 7.90 & 7.30 \\
    \quad OpenAI-DR         & 0.3580 & 0.1260 & 0.2920 & 8.70 & 8.25 & 4.40 & 6.70 & 0.3880 & 0.1360 & 0.2800 & 8.80 & 8.85 & 8.10 & 7.50 \\
    \midrule
    \multicolumn{15}{l}{\textit{Personalized}} \\
    \quad Profile Prompting & 0.3459 & 0.1245 & 0.2789 & 8.52 & 8.14 & 4.49 & 6.70 & 0.3724 & 0.1234 & 0.2641 & 8.72 & 8.75 & 8.24 & 7.62 \\
    \quad Iterative RAG     & 0.3579 & 0.1274 & 0.2877 & 8.90 & 8.37 & \underline{4.78} & \underline{6.74} & 0.3813 & 0.1357 & 0.2815 & \underline{9.09} & \underline{9.03} & \underline{8.30} & \textbf{9.71} \\
    \quad \textbf{PDR (Ours)} & 0.3684 & 0.1293 & \underline{0.3050} & \underline{9.04} & 8.44 & \textbf{6.89} & \textbf{7.20} & 0.3884 & 0.1344 & 0.2853 & 9.05 & 8.83 & \textbf{9.20} & \underline{9.00} \\
    \bottomrule
  \end{tabular}
  }
\end{table*}

\begin{figure}
    \centering
    \includegraphics[width=0.95\linewidth]{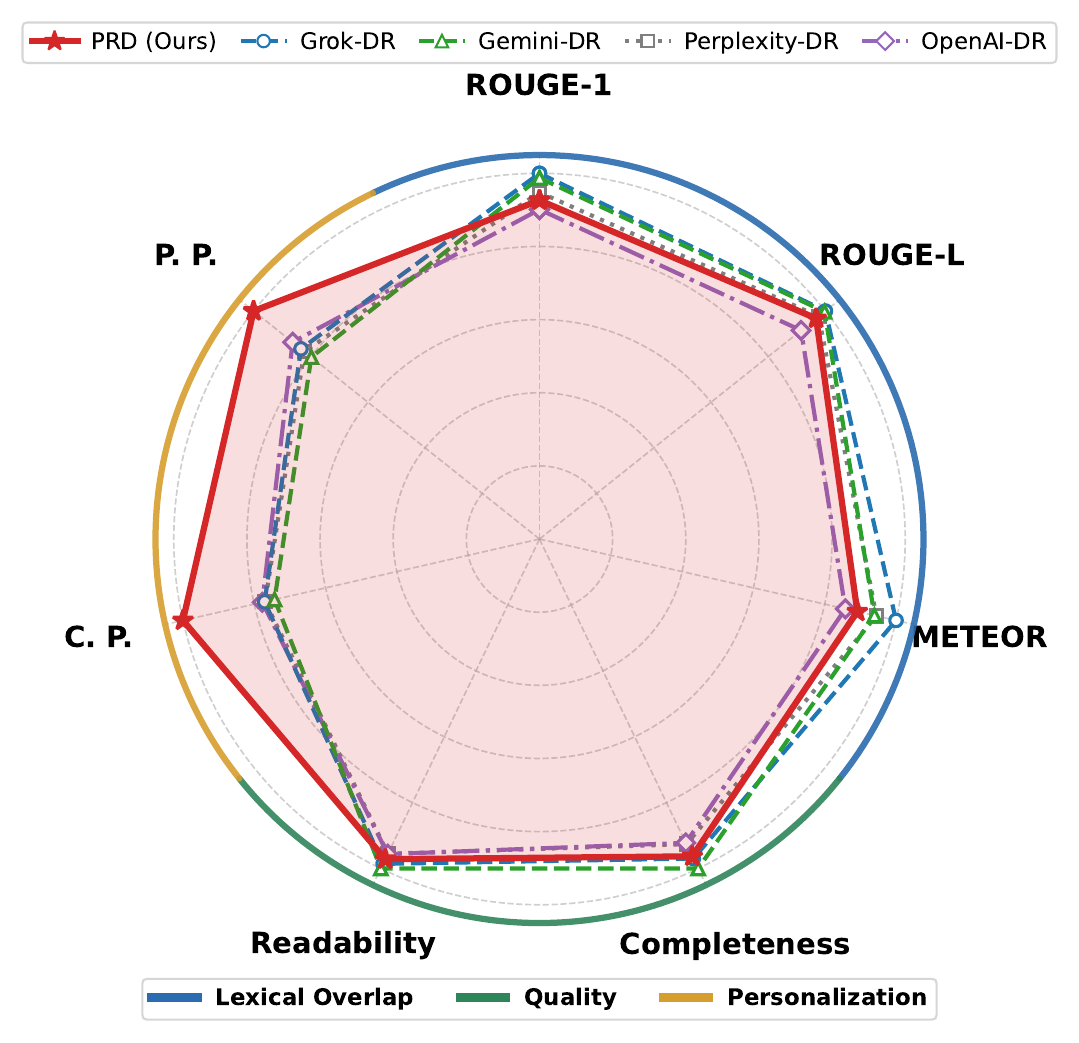}
    \caption{Overall performance comparison with Deep Research Agents. PDR achieves superior Personalization (C.P., P.P.) while maintaining competitive scores in Quality and Lexical metrics.   
    }
    \label{fig:DR_compare}
\end{figure}

\subsection{Experiment Details}
Implementation Details. We constructed a local private retrieval system using a Milvus~\cite{wang2021milvus} vector database paired with BGE-M3~\cite{chen2024bge} embeddings. To ensure experimental reproducibility, we utilized Wikipedia-18 as the external knowledge source—adopting the protocol established by FlashRAG~\cite{jin2025flashrag}—rather than employing non-deterministic real-time web retrieval. For the pipeline's core reasoning capabilities, we selected DeepSeek R1-671B~\cite{deepseek2025r1} as the base models. DeepSeek R1-671B also serves as the evaluation judge to assess system performance.

\subsection{Baselines}
To rigorously evaluate our proposed method given the absence of direct prior work in personalized deep research, we establish two categories of baselines. First, we implement four methodological baselines to assess component contributions: (i) \textit{Zero-shot}; (ii) \textit{+Search}, which uses the input query and includes evidence retrieved from public web search; (iii) \textit{Profile Prompting}, which injects structured user context into the system prompt based on ``+search'' version; and (iv) \textit{Iterative RAG}~\cite{asai2024self}, which exclusively employs a private search agent for multi-step retrieval. Second, we benchmark against four industrial deep research systems—Grok, Perplexity, Gemini, and OpenAI Deep Research—to situate our performance within the current landscape.

\subsection{Result Analysis}
The overall comparison with different Deep Research agents is presented in Figure~\ref{fig:DR_compare}, and the results for each dataset are detailed in Table~\ref{tab:R1_performance_compact}. Our analysis is provided below.
\begin{itemize}[leftmargin=*]
    \item Analysis of the Deep Research Agents (Grok-DR, Perplexity-DR, and Gemini-DR) reveals a distinct trend: these models consistently secure the highest scores in content-centric metrics, such as ROUGE and quality metrics. For instance, Gemini-DR leads in Task 1 with a ROUGE-1 score of 0.3485 and a comprehensiveness score of 6.10, while Perplexity-DR dominates Task 3 and Task 4 in terms of ROUGE metrics. However, this general proficiency does not effectively translate to personalized metrics, specifically Contextual Personalization (C.P.) and Presentation Personalization (P.P.). This limitation arises because these agents primarily rely on external web knowledge, which hinders their ability to access and incorporate the specific personalized information required for these tasks.
    \item In contrast, our proposed PDR framework achieves state-of-the-art performance in personalization capabilities while maintaining competitive content quality. For example, in Task 1, PDR obtains a C.P. score of 7.82 and a P.P. score of 9.51, substantially outperforming the leading general-purpose agent (OpenAI-DR), which scores 5.40 and 7.55, respectively. This superior performance is attributed to two main factors: the highly personalized nature of our dataset and our pipeline design, which ensures the continuous injection of personalized information throughout the entire workflow, from extraction to retrieval.
\end{itemize}

%% file: 2Relatedwork.tex
\section{Related work}

Deep Research Agents represent a significant evolution from traditional RAG systems by embedding autonomous agents capable of complex reasoning, tool orchestration, and reflection \cite{singh2025agentic, zhang2026evoking}. The feasibility of such multi-agent architectures for sophisticated information retrieval has been demonstrated by commercial systems like OpenAI~\cite{openai_deep_research_2025}, Google Gemini~\cite{gemini_deep_research_2024}, and Perplexity~\cite{perplexity_deep_research_2025}. Currently, these deep research pipelines follow two primary methodological approaches: the first emphasizes explicit multi-agent collaboration, where distinct agents handle planning, question formulation, and tool calling within a coordinated workflow to optimize performance \cite{anthropic_multi_agent_2025}; the second employs reinforcement learning (RL) based optimization, exemplified by Kimi-researcher's use of on-policy training and outcome rewards \cite{moonshot_kimi_researcher_2025}. Recent work further examines the trade-off between process-level and outcome-level rewards in agentic RAG settings~\cite{zhang2025reasonrag} and proposes causal intervention methods to align the decision boundaries of deep search agents~\cite{zhang2026tosearch}. However, these architectures remain largely domain-agnostic. Although personalization has long been recognized as an important problem~\cite{zhang2026personalize, li2023hamur, li2025mta, li2025survey_1, gao2025llm4rerank, fu2025unified, liu2025llmemb, wang2023plate, liu2025large} and has been explored in RAG systems, existing efforts remain fragmented across isolated pipeline stages. These include query reformulation with demographic personas~\cite{li2023agent4ranking}, zero-shot query expansion~\cite{jia2024mill}, and production-level rewriting~\cite{microsoft2024}; retrieval methods based on memory-augmented reasoning~\cite{salemi2025rest} and adaptive multi-aspect retrieval augmentation~\cite{xu2025amar}; and generation-stage approaches that use token-level rewards, style transfer~\cite{chen2024pad}, or LLM-powered user simulation~\cite{zhang2025llmsimulator}. Although recent surveys identify a structural convergence between personalized RAG and agentic architectures \cite{li2025survey}, current works lack integration; our framework addresses this gap by maintaining a coherent user model throughout the entire pipeline, enabling holistic personalization that adapts from task planning through to final generation.

%% file: 7Conclusion.tex
\section{Conclusion}
We addressed the limitations of generic ``one-size-fits-all'' Deep Research systems by introducing Personalized Deep Research (PDR). PDR systematically integrates user context via four key components: profile extraction, personalized intent understanding, dynamic dual-stage retrieval, and tailored report generation. Experiments demonstrate that PDR consistently outperforms non personalized baselines and surpasses commercial platforms in personalization capabilities while maintaining competitive quality. To facilitate future research, we introduce PDR-Dataset, the first benchmark for personalized deep research, and PDR-Eval, a multi-dimensional evaluation methodology combining lexical metrics with LLM-as-Judge assessment. This work re-frames Deep Research from a purely technical challenge to a human-centered design problem, establishing a foundation for research assistants that are both accurate and adaptive to individual needs.

\begin{acks}
This research was partially supported by National Natural Science Foundation of China (No.62502404), Hong Kong Research Grants Council (Research Impact Fund No.R1015-23, Collaborative Research Fund No.C1043-24GF, General Research Fund No. 11218325), Institute of Digital Medicine of City University of Hong Kong (No.9229503), and Huawei (Huawei Innovation Research Program).
\end{acks}

%% file: bibfile.bib
@article{xu2025researcherbench,
  title={ResearcherBench: Evaluating Deep AI Research Systems on the Frontiers of Scientific Inquiry},
  author={Xu, Tianze and Lu, Pengrui and Ye, Lyumanshan and Hu, Xiangkun and Liu, Pengfei},
  journal={arXiv preprint arXiv:2507.16280},
  year={2025}
}

@article{du2025deepresearch,
  title={DeepResearch Bench: A Comprehensive Benchmark for Deep Research Agents},
  author={Du, Mingxuan and Xu, Benfeng and Zhu, Chiwei and Wang, Xiaorui and Mao, Zhendong},
  journal={arXiv preprint arXiv:2506.11763},
  year={2025}
}

@inproceedings{tang2008arnetminer,
  title={Arnetminer: extraction and mining of academic social networks},
  author={Tang, Jie and Zhang, Jing and Yao, Limin and Li, Juanzi and Zhang, Li and Su, Zhong},
  booktitle={Proceedings of the 14th ACM SIGKDD international conference on Knowledge discovery and data mining},
  pages={990--998},
  year={2008}
}

@article{kumar2024longlamp,
  title={Longlamp: A benchmark for personalized long-form text generation},
  author={Kumar, Ishita and Viswanathan, Snigdha and Yerra, Sushrita and Salemi, Alireza and Rossi, Ryan A and Dernoncourt, Franck and Deilamsalehy, Hanieh and Chen, Xiang and Zhang, Ruiyi and Agarwal, Shubham and others},
  journal={arXiv preprint arXiv:2407.11016},
  year={2024}
}

@inproceedings{volske2017tl,
  title={Tl; dr: Mining reddit to learn automatic summarization},
  author={V{\"o}lske, Michael and Potthast, Martin and Syed, Shahbaz and Stein, Benno},
  booktitle={Proceedings of the Workshop on New Frontiers in Summarization},
  pages={59--63},
  year={2017}
}

@article{singh2025agentic,
  title={Agentic retrieval-augmented generation: A survey on agentic rag},
  author={Singh, Aditi and Ehtesham, Abul and Kumar, Saket and Khoei, Tala Talaei},
  journal={arXiv preprint arXiv:2501.09136},
  year={2025}
}

@article{li2023agent4ranking,
  title={Agent4ranking: Semantic robust ranking via personalized query rewriting using multi-agent llm},
  author={Li, Xiaopeng and Su, Lixin and Jia, Pengyue and Zhao, Xiangyu and Cheng, Suqi and Wang, Junfeng and Yin, Dawei},
  journal={arXiv preprint arXiv:2312.15450},
  year={2023}
}

@article{li2025survey,
  title={A survey of personalization: From rag to agent},
  author={Li, Xiaopeng and Jia, Pengyue and Xu, Derong and Wen, Yi and Zhang, Yingyi and Zhang, Wenlin and Wang, Wanyu and Wang, Yichao and Du, Zhaocheng and Li, Xiangyang and others},
  journal={arXiv preprint arXiv:2504.10147},
  year={2025}
}

@misc{microsoft2024,
  author       = {Berntson, Alec and Stoica Beck, Alina and Salvador Aguilera, Amaia and Sunavala, Farzad and Gisselbrecht, Thibault and Chen, Xianshun},
  title        = {Raising the bar for RAG excellence: query rewriting and new semantic ranker},
  howpublished = {Microsoft Azure AI Services Blog},
  year         = {2024},
  month        = {nov},
  day          = {19},
  note         = {Announcing generative query rewriting and next‑gen semantic ranker in Azure AI Search},
}

@article{chen2024pad,
  title={Pad: Personalized alignment at decoding-time},
  author={Chen, Ruizhe and Zhang, Xiaotian and Luo, Meng and Chai, Wenhao and Liu, Zuozhu},
  journal={arXiv e-prints},
  pages={arXiv--2410},
  year={2024}
}

@article{salemi2025rest,
  title={Reasoning-Enhanced Self-Training for Long-Form Personalized Text Generation},
  author={Salemi, Alireza and Li, Cheng and Zhang, Mingyang and Mei, Qiaozhu and Kong, Weize and Chen, Tao and Li, Zhuowan and Bendersky, Michael and Zamani, Hamed},
  journal={arXiv preprint arXiv:2501.04167},
  year={2025}
}

@article{li2025webthinker,
  title={Webthinker: Empowering large reasoning models with deep research capability},
  author={Li, Xiaoxi and Jin, Jiajie and Dong, Guanting and Qian, Hongjin and Zhu, Yutao and Wu, Yongkang and Wen, Ji-Rong and Dou, Zhicheng},
  journal={arXiv preprint arXiv:2504.21776},
  year={2025}
}

@article{zheng2025deepresearcher,
  title={Deepresearcher: Scaling deep research via reinforcement learning in real-world environments},
  author={Zheng, Yuxiang and Fu, Dayuan and Hu, Xiangkun and Cai, Xiaojie and Ye, Lyumanshan and Lu, Pengrui and Liu, Pengfei},
  journal={arXiv preprint arXiv:2504.03160},
  year={2025}
}

@article{schmidgall2025agent,
  title={Agent laboratory: Using llm agents as research assistants},
  author={Schmidgall, Samuel and Su, Yusheng and Wang, Ze and Sun, Ximeng and Wu, Jialian and Yu, Xiaodong and Liu, Jiang and Liu, Zicheng and Barsoum, Emad},
  journal={arXiv preprint arXiv:2501.04227},
  year={2025}
}

@article{tang2025ai,
  title={AI-Researcher: Autonomous Scientific Innovation},
  author={Tang, Jiabin and Xia, Lianghao and Li, Zhonghang and Huang, Chao},
  journal={arXiv preprint arXiv:2505.18705},
  year={2025}
}

@online{openai_deep_research_2025,
  author    = {{OpenAI}},
  title     = {Introducing Deep Research},
  year      = {2025},
  date      = {2025-02-02},
  url       = {https://openai.com/zh-Hans-CN/index/introducing-deep-research/},
  urldate   = {2025-08-01},
  language  = {zh-CN}
}

@online{perplexity_deep_research_2025,
  author    = {{Perplexity Team}},
  title     = {Introducing Perplexity Deep Research},
  year      = {2025},
  date      = {2025-02-14},
  url       = {https://www.perplexity.ai/hub/blog/introducing-perplexity-deep-research},
  urldate   = {2025-08-01},
  language  = {en}
}

@online{gemini_deep_research_2024,
  author    = {{Google}},
  title     = {Gemini Deep Research — your personal research assistant},
  year      = {2024},
  date      = {2024-12-11},
  url       = {https://gemini.google/overview/deep-research/},
  urldate   = {2025-08-01},
  language  = {en}
}

@article{phan2025humanity,
  title={Humanity's last exam},
  author={Phan, Long and Gatti, Alice and Han, Ziwen and Li, Nathaniel and Hu, Josephina and Zhang, Hugh and Zhang, Chen Bo Calvin and Shaaban, Mohamed and Ling, John and Shi, Sean and others},
  journal={arXiv preprint arXiv:2501.14249},
  year={2025}
}

@article{wei2024measuring,
  title={Measuring short-form factuality in large language models},
  author={Wei, Jason and Karina, Nguyen and Chung, Hyung Won and Jiao, Yunxin Joy and Papay, Spencer and Glaese, Amelia and Schulman, John and Fedus, William},
  journal={arXiv preprint arXiv:2411.04368},
  year={2024}
}

@inproceedings{mialon2024gaia,
  title={{GAIA}: a benchmark for General {AI} Assistants},
  author={Gr{\'e}goire Mialon and Cl{\'e}mentine Fourrier and Thomas Wolf and Yann LeCun and Thomas Scialom},
  booktitle={The Twelfth International Conference on Learning Representations},
  year={2024},
  url={https://openreview.net/forum?id=fibxvahvs3}
}

@online{anthropic_multi_agent_2025,
  author      = {Hadfield, Jeremy and Zhang, Barry and Lien, Kenneth and Scholz, Florian and Fox, Jeremy and Ford, Daniel},
  title       = {How we built our multi-agent research system},
  organization= {Anthropic PBC},
  year        = {2025},
  date        = {2025-06-13},
  url         = {https://www.anthropic.com/engineering/built-multi-agent-research-system},
  language    = {en}
}

@online{moonshot_kimi_researcher_2025,
  author    = {{Moonshot AI}},
  title     = {Kimi-Researcher: End-to-End RL Training for Emerging Agentic Capabilities},
  year      = {2025},
  date      = {2025-06-20},
  url       = {https://moonshotai.github.io/Kimi-Researcher/},
  language  = {en}
}

@online{substack2025,
  author    = {Author's Name},
  title     = {Title of the Article},
  year      = {2025},
  month     = {aug},
  url       = {https://example.substack.com/p/article-title},
  note      = {Accessed: 2025-08-01}
}

@misc{MediumWebsite,
  author       = {{Medium}},
  title        = {Medium},
  howpublished = {\url{https://medium.com}},
  note         = {Accessed: 2025‑08‑01},
  year         = {2012}
}

@misc{TEDWebsite,
  author       = {{TED Conferences, LLC}},
  title        = {TED},
  howpublished = {\url{https://ted.com}},
  note         = {Accessed: 2025‑08‑01},
  year         = {1984}
}

@inproceedings{banerjee2005meteor,
  title={METEOR: An automatic metric for MT evaluation with improved correlation with human judgments},
  author={Banerjee, Satanjeev and Lavie, Alon},
  booktitle={Proceedings of the acl workshop on intrinsic and extrinsic evaluation measures for machine translation and/or summarization},
  pages={65--72},
  year={2005}
}

@inproceedings{lin-2004-rouge,
    title = "{ROUGE}: A Package for Automatic Evaluation of Summaries",
    author = "Lin, Chin-Yew",
    booktitle = "Text Summarization Branches Out",
    month = jul,
    year = "2004",
    address = "Barcelona, Spain",
    publisher = "Association for Computational Linguistics",
    url = "https://aclanthology.org/W04-1013/",
    pages = "74--81"
}

@inproceedings{wang2021milvus,
  author    = {Jianguo Wang and Xiaomeng Yi and Rentong Guo and Hai Jin and Peng Xu
               and Shengjun Li and Xiangyu Wang and Xiangzhou Guo and Chengming Li
               and Xiaohai Xu and Kun Yu and others},
  title     = {Milvus: A Purpose-Built Vector Data Management System},
  booktitle = {Proceedings of the 2021 International Conference on Management of Data (SIGMOD ’21)},
  year      = {2021},
  doi       = {10.1145/3448016.3457550},
  url       = {https://doi.org/10.1145/3448016.3457550}
}

@article{jin2025flashrag,
  author  = {Jiajie Jin and Yutao Zhu and Guanting Dong and Yuyao Zhang and Xinyu Yang
             and Chenghao Zhang and Tong Zhao and Zhao Yang and Zhicheng Dou and Ji-Rong Wen},
  title   = {FlashRAG: A Modular Toolkit for Efficient Retrieval-Augmented Generation Research},
  journal = {arXiv preprint arXiv:2405.13576},
  note    = {Resource track, WWW 2025 (to appear)},
  year    = {2025},
  url     = {https://arxiv.org/abs/2405.13576}
}

@article{deepseek2025r1,
  author  = {{DeepSeek-AI} and Daya Guo and Dejian Yang and Haowei Zhang and Junxiao Song and Ruoyu Zhang and others},
  title   = {DeepSeek-R1: Incentivizing Reasoning Capability in LLMs via Reinforcement Learning},
  journal = {arXiv preprint arXiv:2501.12948},
  year    = {2025},
  url     = {https://arxiv.org/abs/2501.12948}
}

@article{chen2024bge,
  title={Bge m3-embedding: Multi-lingual, multi-functionality, multi-granularity text embeddings through self-knowledge distillation},
  author={Chen, Jianlv and Xiao, Shitao and Zhang, Peitian and Luo, Kun and Lian, Defu and Liu, Zheng},
  journal={arXiv preprint arXiv:2402.03216},
  year={2024}
}

@article{asai2024self,
  title={Self-rag: Learning to retrieve, generate, and critique through self-reflection},
  author={Asai, Akari and Wu, Zeqiu and Wang, Yizhong and Sil, Avirup and Hajishirzi, Hannaneh},
  year={2024},
  publisher={ICLR}
}

@article{zhang2025survey_dr,
  title={Deep research: A survey of autonomous research agents},
  author={Zhang, Wenlin and Li, Xiaopeng and Zhang, Yingyi and Jia, Pengyue and Wang, Yichao and Guo, Huifeng and Liu, Yong and Zhao, Xiangyu},
  journal={arXiv preprint arXiv:2508.12752},
  year={2025}
}

@inproceedings{zhang2026tosearch,
  title={To Search or Not to Search: Aligning the Decision Boundary of Deep Search Agents via Causal Intervention},
  author={Zhang, Wenlin and Dong, Kuicai and Li, Junyi and Zhang, Yingyi and Li, Xiaopeng and Jia, Pengyue and Wen, Yi and Xu, Derong and Wang, Maolin and Wang, Yichao and others},
  booktitle={Proceedings of the ACM Web Conference 2026},
  pages={2049--2059},
  year={2026}
}

@article{zhang2025reasonrag,
  title={Process vs. outcome reward: Which is better for agentic RAG reinforcement learning},
  author={Zhang, Wenlin and Li, Xiangyang and Dong, Kuicai and Wang, Yichao and Jia, Pengyue and Li, Xiaopeng and Zhang, Yingyi and Xu, Derong and Du, Zhaocheng and Guo, Huifeng and others},
  journal={arXiv preprint arXiv:2505.14069},
  year={2025}
}

@article{li2025aisearch,
  title={Towards ai search paradigm},
  author={Li, Yuchen and Cai, Hengyi and Kong, Rui and Chen, Xinran and Chen, Jiamin and Yang, Jun and Zhang, Haojie and Li, Jiayi and Wu, Jiayi and Chen, Yiqun and others},
  journal={arXiv preprint arXiv:2506.17188},
  year={2025}
}

@inproceedings{xu2025amar,
  title={Harnessing large language models for knowledge graph question answering via adaptive multi-aspect retrieval-augmentation},
  author={Xu, Derong and Li, Xinhang and Zhang, Ziheng and Lin, Zhenxi and Zhu, Zhihong and Zheng, Zhi and Wu, Xian and Zhao, Xiangyu and Xu, Tong and Chen, Enhong},
  booktitle={Proceedings of the AAAI Conference on Artificial Intelligence},
  volume={39},
  number={24},
  pages={25570--25578},
  year={2025}
}

@inproceedings{jia2024mill,
  title={Mill: Mutual verification with large language models for zero-shot query expansion},
  author={Jia, Pengyue and Liu, Yiding and Zhao, Xiangyu and Li, Xiaopeng and Hao, Changying and Wang, Shuaiqiang and Yin, Dawei},
  booktitle={Proceedings of the 2024 Conference of the North American Chapter of the Association for Computational Linguistics: Human Language Technologies (Volume 1: Long Papers)},
  pages={2498--2518},
  year={2024}
}

@article{zhao2018drl,
  title={" Deep reinforcement learning for search, recommendation, and online advertising: a survey" by Xiangyu Zhao, Long Xia, Jiliang Tang, and Dawei Yin with Martin Vesely as coordinator},
  author={Zhao, Xiangyu and Xia, Long and Tang, Jiliang and Yin, Dawei},
  journal={ACM sigweb newsletter},
  volume={2019},
  number={Spring},
  pages={1--15},
  year={2019},
  publisher={ACM New York, NY, USA}
}

@inproceedings{zhang2025llmsimulator,
  title={Llm-powered user simulator for recommender system},
  author={Zhang, Zijian and Liu, Shuchang and Liu, Ziru and Zhong, Rui and Cai, Qingpeng and Zhao, Xiangyu and Zhang, Chunxu and Liu, Qidong and Jiang, Peng},
  booktitle={Proceedings of the AAAI Conference on Artificial Intelligence},
  volume={39},
  number={12},
  pages={13339--13347},
  year={2025}
}

@inproceedings{zhang2026evoking,
title={Evoking User Memory: Personalizing {LLM} via Recollection-Familiarity Adaptive Retrieval},
author={Yingyi Zhang and Junyi Li and Wenlin Zhang and Pengyue Jia and Xianneng Li and Yichao Wang and Derong Xu and Yi Wen and Huifeng Guo and Yong Liu and Xiangyu Zhao},
booktitle={The Fourteenth International Conference on Learning Representations},
year={2026},
url={https://openreview.net/forum?id=f7p0F2X6XN}
}

@inproceedings{li2023hamur,
  title={Hamur: Hyper adapter for multi-domain recommendation},
  author={Li, Xiaopeng and Yan, Fan and Zhao, Xiangyu and Wang, Yichao and Chen, Bo and Guo, Huifeng and Tang, Ruiming},
  booktitle={Proceedings of the 32nd ACM International Conference on Information and Knowledge Management},
  pages={1268--1277},
  year={2023}
}

@article{li2025mta,
  title={MTA: A Merge-then-Adapt Framework for Personalized Large Language Model},
  author={Li, Xiaopeng and Zheng, Yuanjin and Wang, Wanyu and Jia, Pengyue and Wang, Yiqi and Wang, Maolin and Wei, Xuetao and Zhao, Xiangyu and others},
  journal={arXiv preprint arXiv:2511.20072},
  year={2025}
}

@article{li2025survey_1,
  title={A survey of generative recommendation from a tri-decoupled perspective: Tokenization, architecture, and optimization},
  author={Li, Xiaopeng and Chen, Bo and She, Junda and Cao, Shiteng and Wang, You and Jia, Qinlin and He, Haiying and Zhou, Zheli and Liu, Zhao and Liu, Ji and others},
  year={2025},
  publisher={Preprints}
}

@inproceedings{zhang2026personalize,
  title={Personalize before retrieve: Llm-based personalized query expansion for user-centric retrieval},
  author={Zhang, Yingyi and Jia, Pengyue and Xu, Derong and Wen, Yi and Li, Xianneng and Wang, Yichao and Zhang, Wenlin and Li, Xiaopeng and Gan, Weinan and Guo, Huifeng and others},
  booktitle={Proceedings of the AAAI Conference on Artificial Intelligence},
  volume={40},
  number={19},
  pages={16406--16414},
  year={2026}
}

@inproceedings{liu2025large,
  title={Large Language Model Enhanced Recommender Systems: Methods, Applications and Trends},
  author={Liu, Qidong and Zhao, Xiangyu and Wang, Yuhao and Wang, Yejing and Zhang, Zijian and Sun, Yuqi and Li, Xiang and Wang, Maolin and Jia, Pengyue and Chen, Chong and others},
  booktitle={Proceedings of the 31st ACM SIGKDD Conference on Knowledge Discovery and Data Mining V. 2},
  pages={6096--6106},
  year={2025}
}

@inproceedings{wang2023plate,
  title={PLATE: A prompt-enhanced paradigm for multi-scenario recommendations},
  author={Wang, Yuhao and Zhao, Xiangyu and Chen, Bo and Liu, Qidong and Guo, Huifeng and Liu, Huanshuo and Wang, Yichao and Zhang, Rui and Tang, Ruiming},
  booktitle={Proceedings of the 46th International ACM SIGIR Conference on Research and Development in Information Retrieval},
  pages={1498--1507},
  year={2023}
}

@inproceedings{liu2025llmemb,
  title={Llmemb: Large language model can be a good embedding generator for sequential recommendation},
  author={Liu, Qidong and Wu, Xian and Wang, Wanyu and Wang, Yejing and Zhu, Yuanshao and Zhao, Xiangyu and Tian, Feng and Zheng, Yefeng},
  booktitle={Proceedings of the AAAI Conference on Artificial Intelligence},
  volume={39},
  number={11},
  pages={12183--12191},
  year={2025}
}

@article{fu2025unified, 
  title={A unified framework for multi-domain ctr prediction via large language models},
  author={Fu, Zichuan and Li, Xiangyang and Wu, Chuhan and Wang, Yichao and Dong, Kuicai and Zhao, Xiangyu and Zhao, Mengchen and Guo, Huifeng and Tang, Ruiming},
  journal={ACM Transactions on Information Systems},
  volume={43},
  number={5},
  pages={1--33},
  year={2025},
  publisher={ACM New York, NY}
}

@inproceedings{gao2025llm4rerank,
  title={Llm4rerank: Llm-based auto-reranking framework for recommendations},
  author={Gao, Jingtong and Chen, Bo and Zhao, Xiangyu and Liu, Weiwen and Li, Xiangyang and Wang, Yichao and Wang, Wanyu and Guo, Huifeng and Tang, Ruiming},
  booktitle={Proceedings of the ACM on Web Conference 2025},
  pages={228--239},
  year={2025}
}
